\documentclass[final,1p,times]{elsarticle} 

\usepackage{graphicx}
\usepackage{amssymb} 
\usepackage{amsthm} 
\usepackage{lineno}
\usepackage{here,array}
\usepackage{wrapft}

\journal{Nuclear Physics A} 

\begin{document}

\begin{frontmatter} 

\title{Heavy quark quenching from RHIC to LHC and the consequences of gluon damping}


\author[auth1]{P.B. Gossiaux} 
\author[auth1]{M. Nahrgang} 
\author[auth1,auth2]{M. Bluhm} 
\author[auth1]{Th. Gousset} 
\author[auth1]{J. Aichelin} 

%

\address[auth1]{SUBATECH UMR 6457 (\'Ecole des Mines de Nantes, IN2P3/CNRS, Université de Nantes), Nantes, France}
\address[auth2]{Dip. di Fisica Teorica, Universit\`a di Torino, via Giuria 1, I-10125 Torino, Italy; INFN, Sezione di Torino}


\begin{abstract} 
In this contribution to the Quark Matter 2012 conference, we study whether
energy loss models established for RHIC energies to describe the quenching of 
heavy quarks can be applied at LHC with the same success. We also 
benefit from the larger $p_T$-range accessible at this accelerator to test
the impact of gluon damping on observables such as the nuclear modification
factor. 
\end{abstract} 

\end{frontmatter} 


\section{Introduction}
Recently, we have proposed a microscopic approach for the quenching of heavy quarks (HQ) in ultra relativistic 
heavy ions collisions (URHIC) \cite{Gossiaux:2008,Gossiaux:2010}, assuming interactions with light partons through both elastic and 
radiative processes evaluated by resorting to some parameterization of the running 
coupling constant, while those partons are spatially distributed along hydrodynamical evolution~\cite{KolbHeinz} of the 
quark gluon plasma (QGP) created in these collisions. This approach was able to explain successfully several observables measured at RHIC, 
such as the nuclear modification factor ($R_{AA}$)and the elliptic flow ($v_2$) of non-photonic single electrons (NPSE). 
The diffusion coefficient $D_s$ for HQ in QGP -- a fundamental property of this state of matter -- could thus be 
extracted \cite{Gossiaux:2011ISMD}. Here, we would like to assess the robustness of our models by confronting their predictions for 
$D$ and $B$ mesons production in URHIC at LHC to some experimental results obtained so far by ALICE and CMS collaborations.

\section{Heavy quark quenching at RHIC}

\begin{figure}[htbp]
\begin{center}
\includegraphics[width=0.45\textwidth]{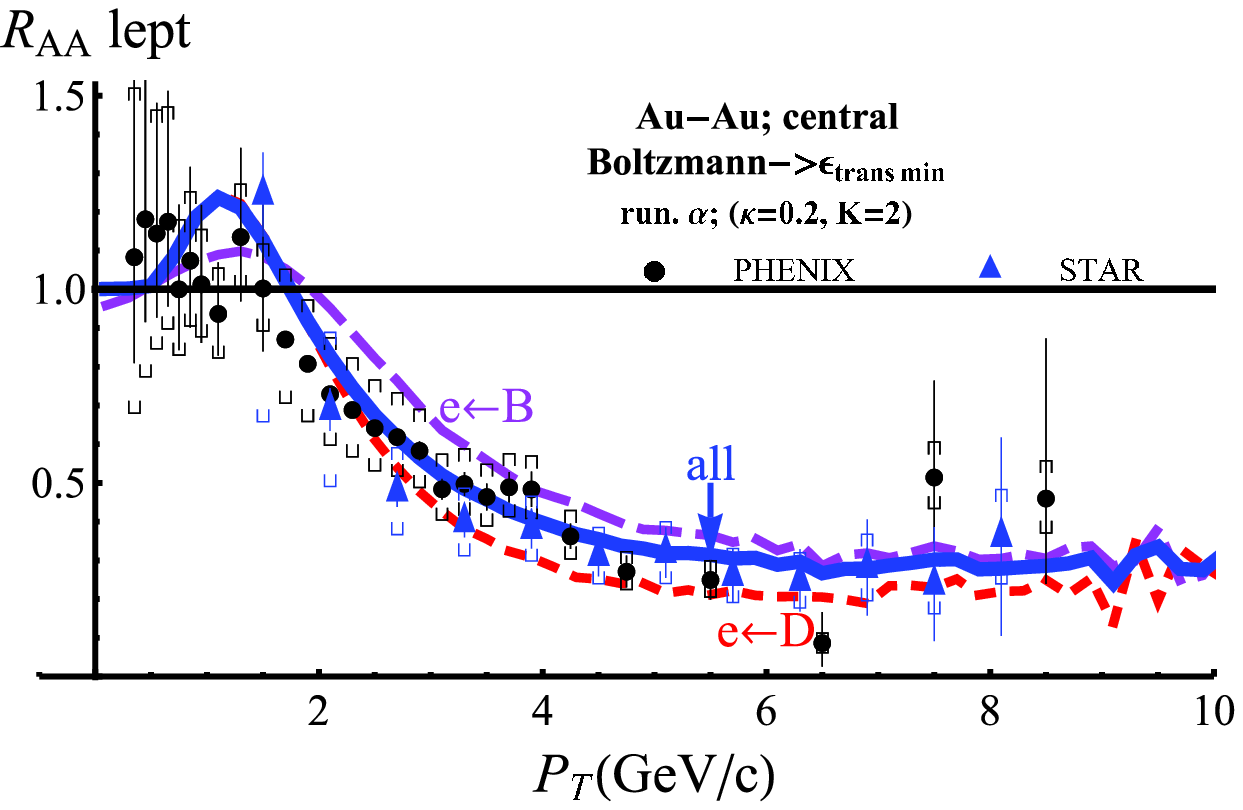}
\hspace{1cm}
\includegraphics[width=0.45\textwidth]{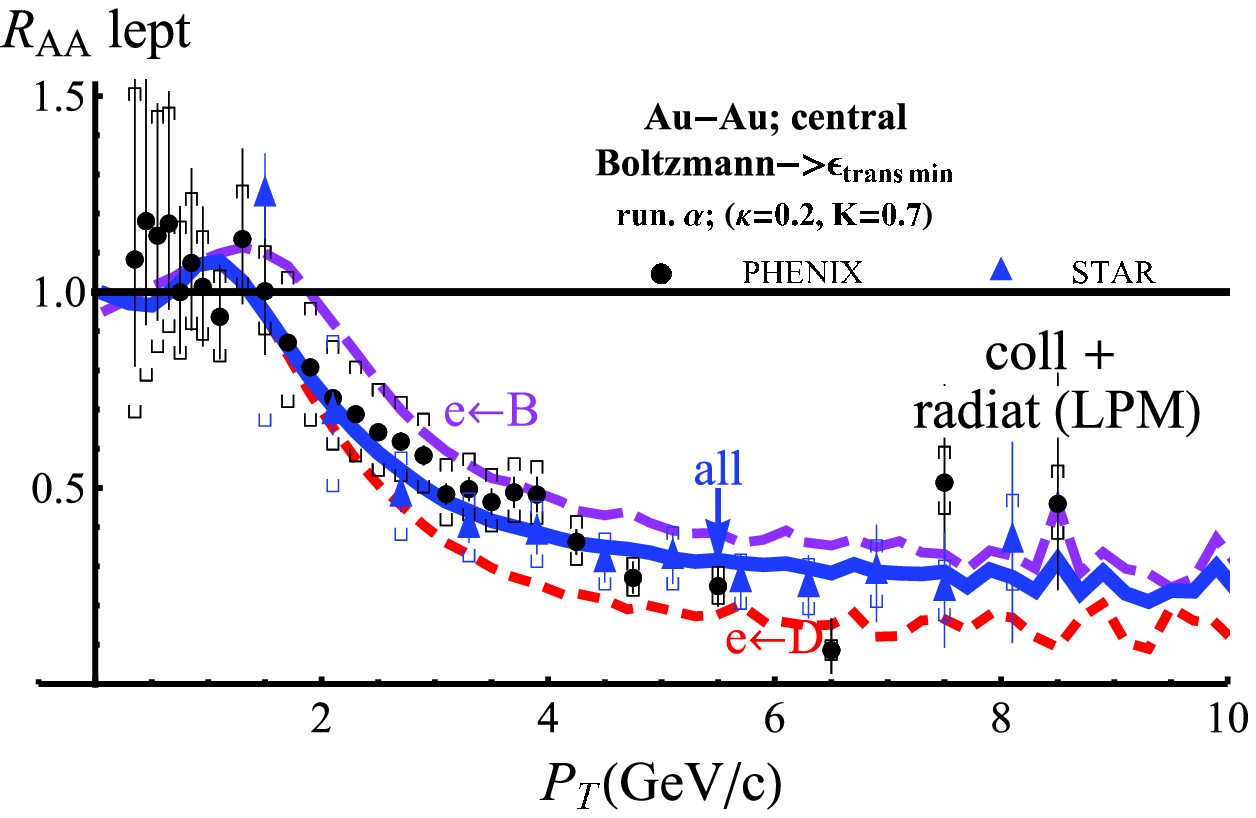}
\end{center}
\caption{Plain: prediction of our elastic ELoss model (left) and of our elastic + radiative (LPM) ELoss model for NPSE $R_{AA}$ 
in central Au-Au URHIC at $\sqrt{s}=200\,{\rm GeV}$. Dashed (dotted): contribution from the $b$ ($c$) quarks.}
\label{fig:RAANPSE}
\end{figure}

Let us recall that our overall strategy is to establish energy loss models based on the interaction rates of HQ with 
the QGP constituents and then allow for some global rescaling by a factor $K$ that mimics the left over ingredients
and the uncertainties affecting the models. In \cite{Gossiaux:2008}, 
 a good agreement was found between our collisional model E and the NPSE observables ($R_{AA}$ for all centralities 
and $v_2$) measured by PHENIX and STAR for $K\approx 2$ (including the mixed phase in the evolution). In \cite{Gossiaux:2010}, 
an equally good agreement for the NPSE was obtained 
with a cocktail of collisional energy loss and radiative energy loss evaluated from a generalization of the Gunion-Bertsch spectrum
\cite{Gunion:1981qs} for heavy quarks. More recently~\cite{Gossiaux:2012HP}, we have considered coherence effects for the radiation, with 
however little consequence on the [0 GeV/c; 10 GeV/c] $p_T$-range presently achievable at RHIC for HQ. Not surprisingly, we still
find a good agreement for the $R_{AA}$ of NPSE with this improved model\footnote{Hereafter referred to as "radiative (LPM)".}, 
for $K\approx 0.7$, as illustrated in fig. \ref{fig:RAANPSE} (right).

\begin{figure}[H]
\begin{center}
\includegraphics[width=0.45\textwidth]{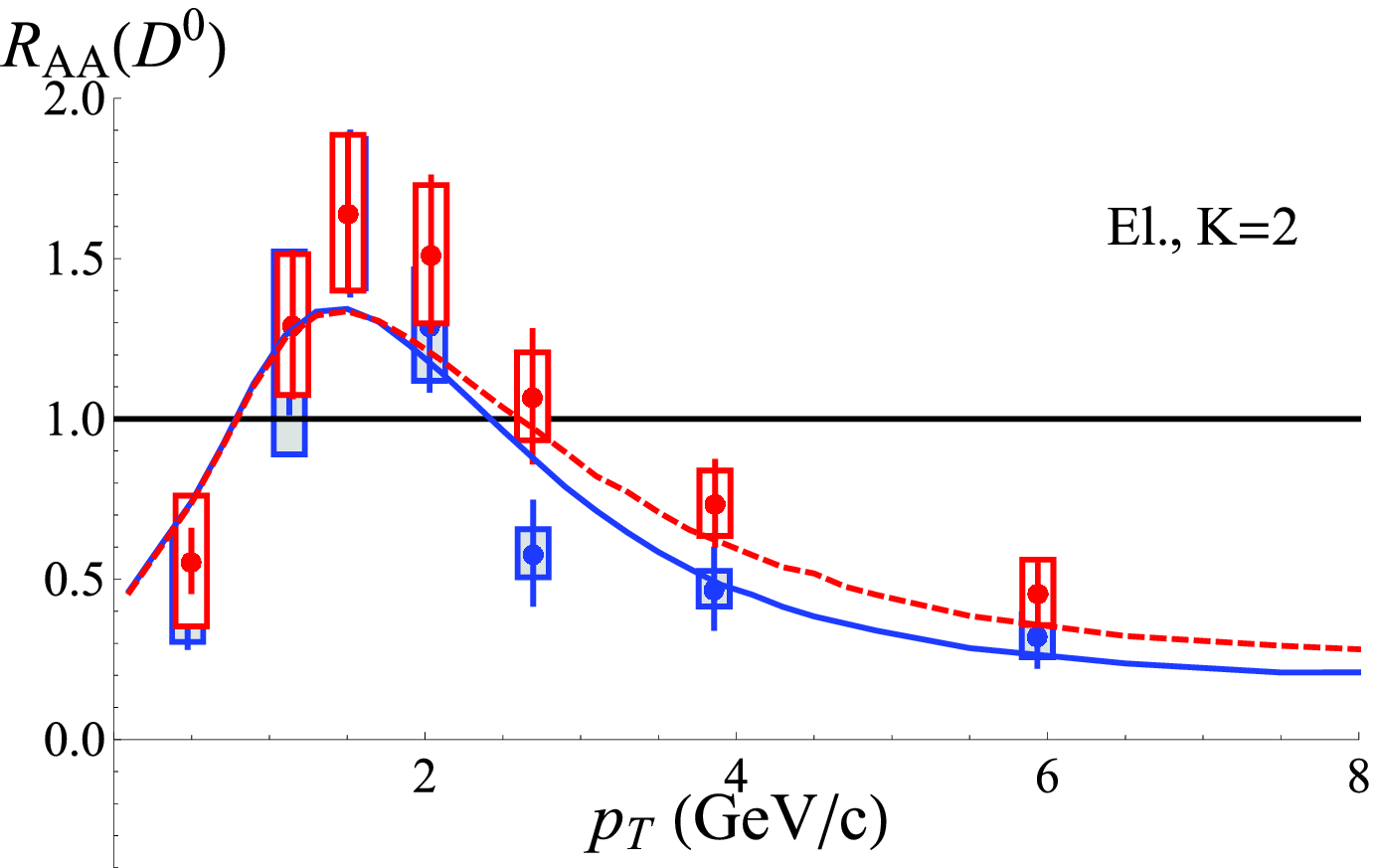}
\hspace{1cm}
\includegraphics[width=0.45\textwidth]{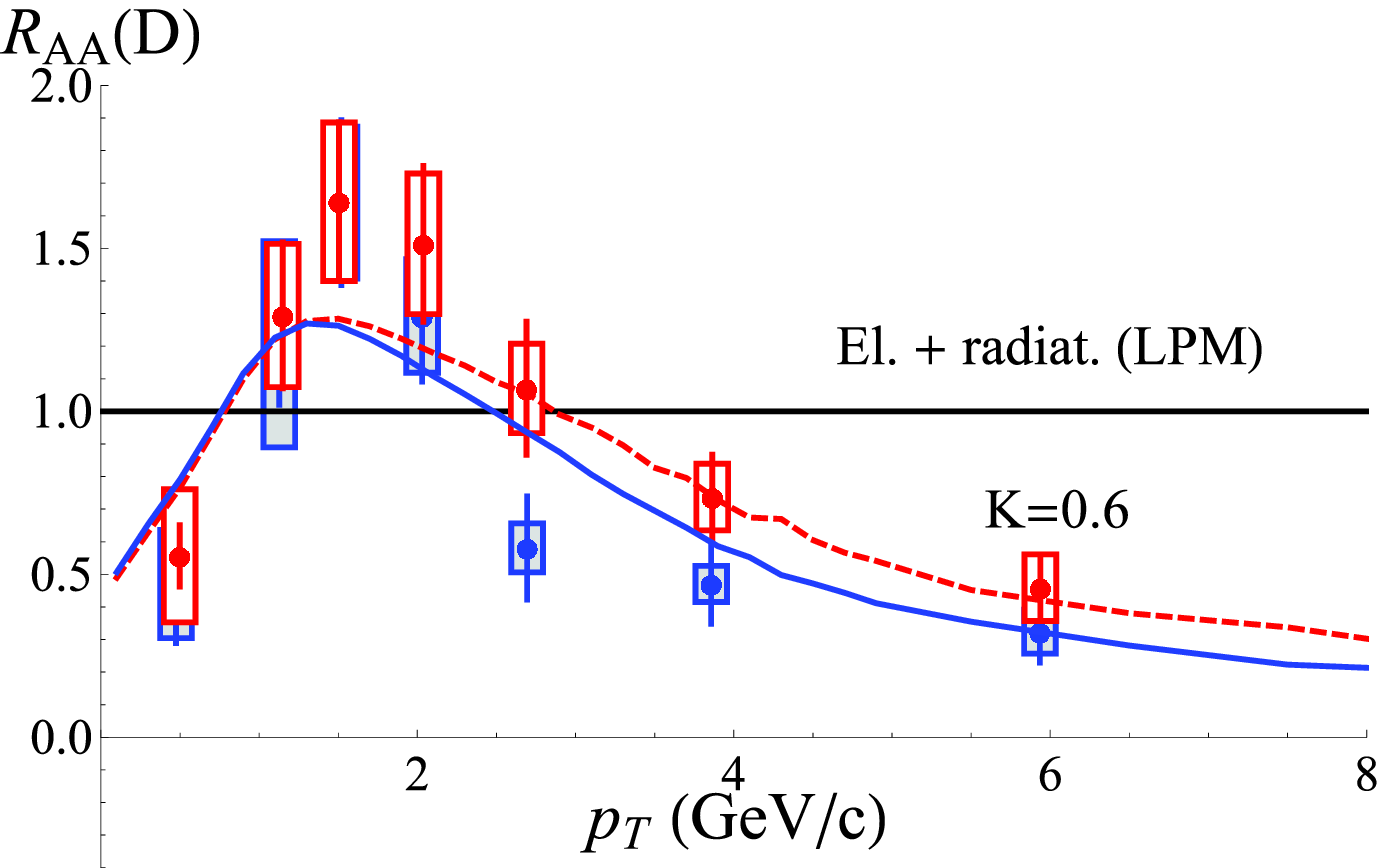}
\end{center}
\caption{Symbols: same as fig. \ref{fig:RAANPSE} for $D^0$ mesons measured by STAR in $0-10\%$ (filled) and $0-80\%$ (open) centrality classes. 
Curves: prediction of our elastic ELoss model (left) and of our elastic + radiative (LPM) ELoss model for $0-10\%$ (plain) 
and $0-80\%$ (dashed) centrality classes.}
\label{fig:RAAD}
\end{figure}

\begin{table}[H]
\begin{center}
\begin{tabular}{|p{0.3cm}|p{0.6cm}|p{1cm}|p{1cm}||p{1cm}|}
\hline
 & \multicolumn{3}{c||}{RHIC} & LHC \\
\hline
\footnotesize $K$ &\footnotesize NPSE & \footnotesize $D^0$ (0-10\%) &
\footnotesize $D^0$ (0-80\%) & \footnotesize $D^0$ (0-20\%) \\
\hline
\footnotesize 1.4 & &  & * & ** \\
\hline
\footnotesize 1.6 & * & * & ** & *** \\
\hline
\footnotesize 1.8 & ** & *** & *** & ** \\
\hline
\footnotesize 2.0 & *** & *** & ** & * \\
\hline
\footnotesize 2.2 & ** & * & * & \\
\hline
\end{tabular}
\begin{tabular}{|p{0.3cm}|p{0.6cm}|p{1cm}|p{1cm}||p{1cm}|}
\hline
 & \multicolumn{3}{c||}{RHIC} & LHC \\
\hline
\footnotesize $K$ &
\footnotesize NPSE & 
\footnotesize $D^0$ (0-10\%) & 
\footnotesize $D^0$ (0-80\%) & 
\footnotesize $D^0$ (0-20\%) \\
\hline
\footnotesize 0.4 &  &  &  & * \\
\hline
\footnotesize 0.5 &  & * & ** & ** \\
\hline
\footnotesize 0.6 & ** & *** & *** & ** \\
\hline
\footnotesize 0.7 & *** & *** & ** & * \\
\hline
\footnotesize 0.8 & ** & * & * & \\
\hline
\end{tabular}
\end{center}
\caption{Optimal values of the $K$ rescaling factor for the collisional energy loss (left) as well as
for the collisional + radiative (LPM) energy loss cocktail (right) for various $R_{AA}$.}
\label{table_D1}
\end{table}

In this contribution, we take the opportunity of the $R_{AA}$ of $D^0$ mesons presented by the STAR 
collaboration~\cite{Xie:2012} for $p_T\in[0\,{\rm GeV/c};6\,{\rm GeV/c}]$ at this conference to better constrain our optimal value of $K$. 
In fig. \ref{fig:RAAD}, we display 
a typical ``best set'' of curves for both models, while in tables \ref{table_D1}, we summarize the best 
$K$-values for the 3 observables considered at RHIC. 
The consistence obtained for both $D^0$ {\em and} NPSE at RHIC is a rather clear indication that the
quenching from $b$ quark at intermediate $p_T$ is correctly described by our models. In fig. 
\ref{fig:RAANPSE}, we show the $R_{AA}$ of leptons stemming independently from $D$ and $B$ mesons. 
As both types of energy loss obey mass hierarchy $\frac{dE(b)}{dx} < \frac{dE(c)}{dx}$, we naturally
find $R_{AA}(e\rightarrow D) < R_{AA}(e\rightarrow B)$ for $p_T\gtrsim 1.5~{\rm GeV}/c$. This is
however in contradiction with the results from the PHENIX collaboration \cite{Nouicer:2012} extracted from 
the fit of distributions of the distance of closest approach and presented at this conference. In our view, 
this puzzle should be clarified at some point by the direct measurement of $B$ mesons.

\section{The LHC case}
\begin{figure}[H]
\begin{center}
\includegraphics[width=0.45\textwidth]{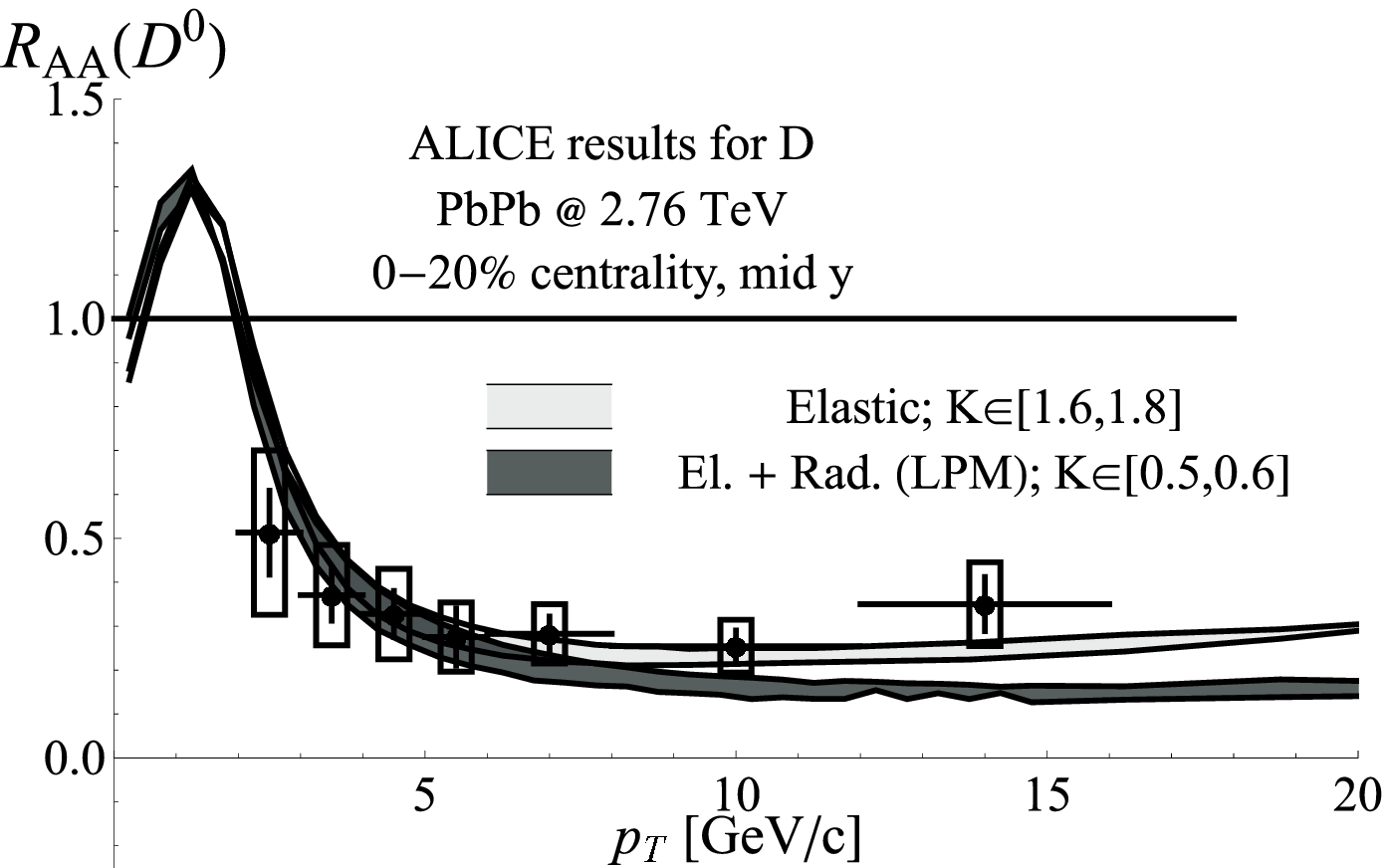}
\includegraphics[width=0.45\textwidth]{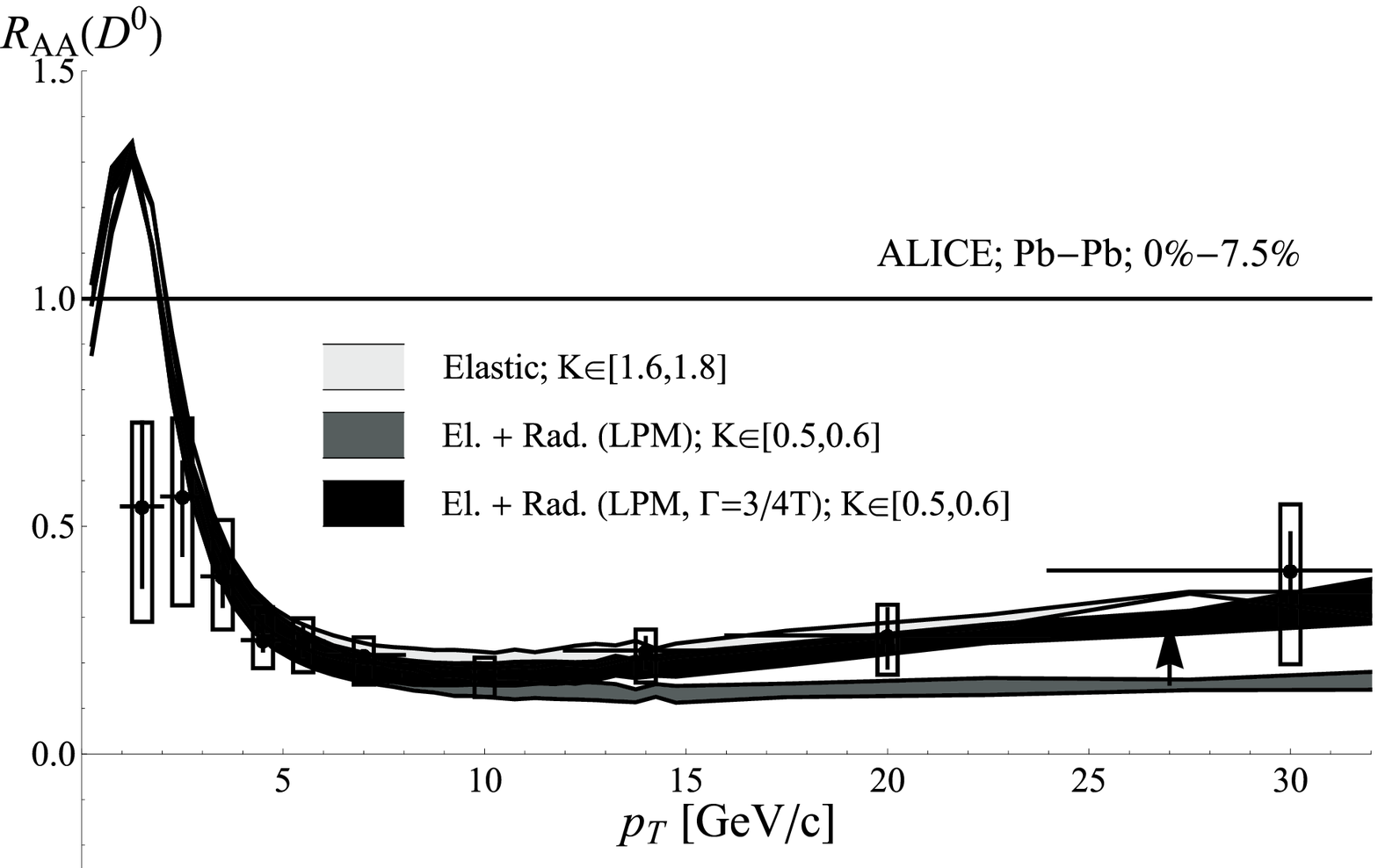}
\end{center}
\caption{Left: $R_{AA}$ of $D$ mesons in 0-20\% centrality Pb-Pb collisions at $\sqrt{s}=2.76~{\rm TeV}$ 
for optimal values
of the rescaling factor $K$; the light gray band represents the elastic energy loss and the gray band the
elastic + radiative (LPM) cocktail. Right: same for the 0-7.5\% centrality class; the dark gray band represents
the impact of gluon damping on the radiative energy loss, evaluated here for an arbitrary width $\Gamma= 0.75 T$.}
\label{fig:RAADLHC}
\end{figure}

We consider the same models for HQ energy loss at LHC, just modifying the initial $p_T$ distribution 
according to the FONLL scheme~\cite{FONLL} as well as the initial entropy density $s_0$ of the QGP phase at the hottest
point in order to reproduce the final density $\frac{dN_{\rm ch}}{dy}=1600$ at mid-rapidity. 
In comparison with the ALICE results~\cite{ALICED:2012}
for $D$ mesons, the optimal $K$ values extracted from RHIC lead to a slight 
excess of quenching~\cite{Gossiaux:2011ISMD,Gossiaux:2012QNP} at intermediate $p_T$ for 
both models, while $v_2(D)$ in good agreement~\cite{SQM2011} with the data. As illustrated in tables
\ref{table_D1} and in fig. \ref{fig:RAADLHC}, a 10\% decrease of the coupling leads to a reasonnable agreement 
for $p_T\lesssim 10~{\rm GeV}/c$. In our mind, this is an acceptable rescaling in view of the 
moderate sophistication of the models, and we would thus argue to have developed a consistent modeling
of heavy quark quenching ``from RHIC to LHC''.

 For larger $p_T$ ($p_T\gtrsim 10~{\rm GeV}/c$), the data seems to 
indicate a rise of the $R_{AA}$ that is better reproduced by the pure collisional energy loss 
component, thanks to
its logarithmic increase at large momenta. This is rather difficult to accept, as one precisely expects the 
dominance of radiative energy loss in this regime. The same feature appears even more clearly for the
$0-7.5\%$ centrality class (fig. \ref{fig:RAADLHC}, right), where data has been measured up to 
$p_T=30~{\rm GeV}/c$. This triggers our interest in new effects neglected up to now, such as 
the impact of gluon damping on radiative energy loss. In \cite{Bluhm:2011}, we have indeed studied the effect of 
an absorptive medium on standard LPM \cite{LPM} radiation in electrodynamics and have advocated that the large time 
needed for the photon formation in Bremsstrahlung from ultra relativistic charges is not affordable if damping is 
taken into account. Similar effect arises in QCD, as we have recently advocated in~\cite{Bluhm_2012}. 
For concrete implementation in our radiative (LPM) model, we consider the gluon radiation according
to the $\frac{d^2I_{\rm LPM}}{dz d\omega}$ spectrum described in~\cite{Gossiaux:2012HP}
 and then quench this 
radiation with an acceptance probability of ${\rm min}(1,\frac{t_d}{l_f})$ where $t_d=\frac{1}{\Gamma}$ is the 
damping time and $l_f$ is the formation length discussed in~\cite{Bluhm_2012}.
\begin{wrapfigure}{r}{6.cm} 
\includegraphics[width=6 cm]{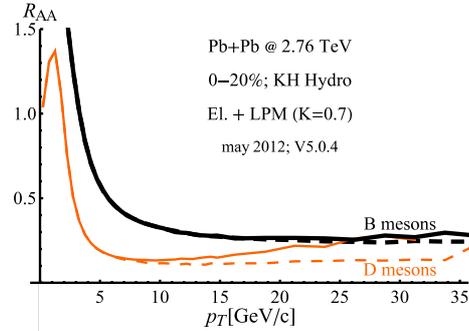}
\caption{$R_{AA}$ of $B$ (thick lines) and $D$ (thin lines) mesons in $0-20\%$ Pb-Pb collisions at $\sqrt{s}=2.76~{\rm TeV}$
with usual el. + radiat. LPM (dashed) as well as including gluon damping (plain) with $\Gamma=0.75 T$.}
\label{dampin_no_damping}
\end{wrapfigure} 
Within pQCD, one obtains \cite{Bluhm_2012, Biro:1993}
that $\Gamma \propto T$. For the purpose of the illustration, we have chosen $\Gamma/T=0.75$ 
and show the consequence of this finite damping in figure \ref{fig:RAADLHC} (right, dark gray band). 
The arrow indicates the shift in the $R_{AA}$ due to gluon damping
which softens the radiation spectra and thus reduces the average energy loss, an effect that manifest itself especially at
larger energies as discussed in \cite{Bluhm_2012}. Although NLO effects such as gluon damping deserve more detailed investigations to be 
performed in the future, let us mention that they can lead to drastic consequences, as for instance the coincidence of both $R_{AA}$
of $D$ and $B$ mesons at rather moderate $p_T$, as illustrated by fig. \ref{dampin_no_damping}. On this fig. it is also 
important to notice that $R_{AA}(B)$ for $p_T\in[6~{\rm GeV}/c;30~{\rm GeV}/c]$ is compatible with the value extracted by the CMS 
collaboration for non-prompt $J/\psi$~\cite{CMS:quarkonia}.

\section{Summary}
In this contribution, we have argued that the effective models of energy loss that we have developed over the past years 
to encompass open heavy flavor observables at RHIC are in pretty good agreement -- within 10\% accuracy -- with 
similar observables at LHC for intermediate $p_T$. At larger $p_T$, new effects neglected up to now might be revealed, as
for instance the damping of high energy gluon radiated in coherent processes.



\begin{thebibliography}{00} 
\bibitem{Gossiaux:2008}
P.B. Gossiaux, J. Aichelin, Phys. Rev. C{\bf 78}, 014904 (2008), [hep-ph/0802.2525].

\bibitem{Gossiaux:2010}
P.B. Gossiaux, V. Guiho, J. Aichelin, J. Phys. G: Nucl. Part. Phys. {\bf 37} (2010) 094019.

\bibitem{KolbHeinz}
P. F. Kolb, J. Sollfrank, and U. Heinz, Phys. Rev. C {\bf 62} (2000) 054909,
P.F. Kolb and U. Heinz, in ``Quark-Gluon Plasma 3'' (World Scientific, Singapore, 2004) [arXiv:nuclth/0305084].

\bibitem{Gossiaux:2011ISMD}
P.B.~Gossiaux, J.~Aichelin and T.~Gousset, 
Prog.\ Th.\ Phys. {\bf 193} (2012) 110 [arXiv:1201.4038].

\bibitem{Gunion:1981qs}
J.~F.~Gunion and G.~Bertsch,
  Phys.\ Rev.\  D {\bf 25} (1982) 746.

\bibitem{Gossiaux:2012HP}
P.B.~Gossiaux, proceedings from the ``Hard probes 2012'' conference [arxiv:1209.0844].

\bibitem{Xie:2012}
Contribution of W.~Xie in this volume.

\bibitem{Nouicer:2012}
R. Nouicer, QM 2012, http://www.phenix.bnl.gov/phenix/WWW/talk/archive/2012/QM12/t1988.ppt.

\bibitem{FONLL}
M.~Cacciari et al., [arXiv:1205.6344].

\bibitem{ALICED:2012}
ALICE collaboration, [arxiv 1203.2160v4].

\bibitem{Gossiaux:2012QNP}
P.B.~Gossiaux et al., proceedings from Sixth International Conference on Quarks and Nuclear Physics [arxiv 1207.5445].

\bibitem{SQM2011}
J.~Aichelin, P.B.~Gossiaux and T.~Gousset, Acta Physica Polonica {\bf 43} (2012) 655 [arXiv:1201.4192v1].

\bibitem{Bluhm:2011}
M. Bluhm, P.B. Gossiaux, and J. Aichelin, arXiv:1106.2856, PRL {\bf 107} (2011) 265004. 

\bibitem{LPM}
L.D. Landau and I. Ya. Pomeranchuk, Dokl. Akad. Nauk  SSSR {\bf 92} (1953) 535; ibid. {\bf 92} (1953) 735.

\bibitem{Bluhm_2012}
M. Bluhm, P. B. Gossiaux, T. Gousset, J. Aichelin, [arXiv:1204.2469v1]. 

\bibitem{Biro:1993}
T. S. Biro et al., Phys. Rev. C {\bf 48} (1993) 1275.

\bibitem{CMS:quarkonia}
CMS collaboration, [arxiv:1201.5069].

\end{thebibliography}
\end{document}